\documentclass[runningheads]{llncs}
\usepackage{graphicx}
\usepackage{amsmath,amssymb} 
\usepackage{color}
 
\usepackage{subfigure}
\usepackage[lined,boxed, ruled]{algorithm2e}
\usepackage{multirow}
\usepackage{diagbox}
\usepackage{wrapfig}
\usepackage{hyperref}
\usepackage{url}
\usepackage[table,xcdraw]{xcolor}
\usepackage{pfnote}

\def\etal{\textit{et al. }}
\newcommand{\alert}[1]{{\textcolor{red}{#1}}}

\newcommand*\samethanks[1][\value{footnote}]{\footnotemark[#1]}

\begin{document}
\pagestyle{headings}
\mainmatter

\title{Encoding Structure-Texture Relation with P-Net for Anomaly Detection in Retinal Images} 
\titlerunning{P-Net for Anomaly Detection}

\author{Kang Zhou\inst{1, \thanks{Kang Zhou and Yuting Xiao equally contribute to this work.}} \and
	Yuting Xiao\inst{1,\samethanks} \and
	Jianlong Yang\inst{2} \and
	Jun Cheng\inst{3}\and
	Wen Liu\inst{1}\and \\
	Weixin Luo\inst{1}\and
	Zaiwang Gu\inst{3}\and
	Jiang Liu\inst{2,4}\and
	Shenghua Gao\inst{1,5,\thanks{Shenghua Gao is the corresponding author.}}}
\authorrunning{Zhou et al.}

\institute{School of Information Science and Technology, ShanghaiTech University, China \\
	\email{\{zhoukang, xiaoyt, liuwen, luowx, gaoshh\}@shanghaitech.edu.cn} \and
Cixi Institute of Biomedical Engineering, Chinese Academy of Sciences, China \\
\email{yangjianlong@nimte.ac.cn} \and
UBTech Research, China \\
\email{juncheng@ieee.org, guzaiwang01@gmail.com} \and
Southern University of Science and Technology, China \\
\email{liuj@sustech.edu.cn}
\and Shanghai Engineering Research Center of Intelligent Vision and Imaging, China
}

\maketitle

\begin{abstract}
Anomaly detection in retinal image refers to the identification of abnormality caused by various retinal diseases/lesions, by only leveraging normal images in training phase. Normal images from healthy subjects often have regular structures (e.g., the structured blood vessels in the fundus image, or structured anatomy in optical coherence tomography image). On the contrary, the diseases and lesions often destroy these structures. Motivated by this, we propose to leverage the relation between the image texture and structure to design a deep neural network for anomaly detection. Specifically, we first extract the structure of the retinal images, then we combine both the structure features and the last layer features extracted from original health image to reconstruct the original input healthy image. The image feature provides the texture information and guarantees the uniqueness of the image recovered from the structure. In the end, we further utilize the reconstructed image to extract the structure and measure the difference between structure extracted from original and the reconstructed image. On the one hand, minimizing the reconstruction difference behaves like a regularizer to guarantee that the image is corrected reconstructed. On the other hand, such structure difference can also be used as a metric for normality measurement. The whole network is termed as P-Net because it has a ``P'' shape. 
Extensive experiments on RESC dataset and iSee dataset validate the effectiveness of our approach for anomaly detection in retinal images. 
Further, our method also generalizes well to novel class discovery in retinal images and anomaly detection in real-world images. 


\end{abstract}


\section{Introduction}
\vspace{-0.20in}

\begin{figure}[htb]
	\centering
	\includegraphics[width=\textwidth]{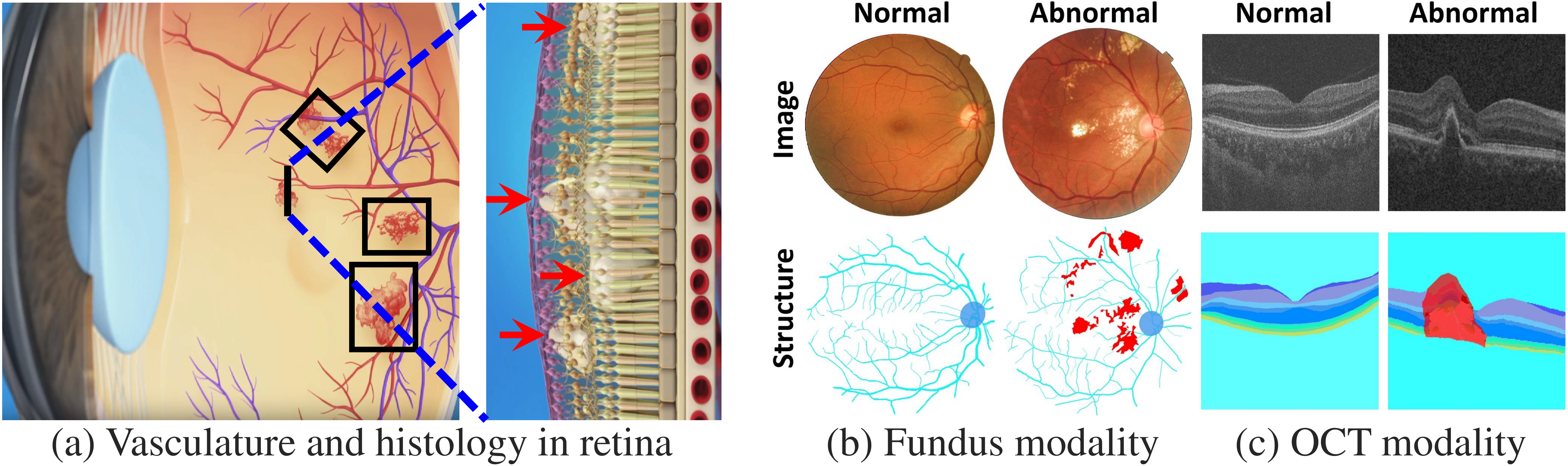}
	\vspace{-0.15in}
	\caption{The motivation of leveraging structure information for anomaly detection. The normal medical images are highly structured, while the regular structure is broken in abnormal images. For example, the lesions (denoted by black bounding box and \alert{red arrow} in (a) of diabetic retinopathy destroy the blood vessel and histology layer in retina. Thus, in the abnormal retinal fundus image and optical coherence tomography (OCT) image, the lesions (denoted by \alert{red color} in (b) and (c)) broke the structure. Moreover, this phenomenon agrees with the cognition of doctors. Motivated by this clinical observation, we suggest utilizing the structure information in anomaly detection. 
	The figure (a) is adopted from the website of American Academy of Ophthalmology \cite{kierstan2019what}.}
	\label{fig:intro_structure}
	\vspace{-0.15in}
\end{figure}

Deep convolutional neural networks (CNNs) have achieved many breakthroughs in medical image analysis \cite{litjens2017survey}\cite{xing2017deep}\cite{zhou2017deep}\cite{zhou2018multi}\cite{fu2018joint}\cite{zhang2019attention}.
However, these methods usually depend on large-scale balanced data in medical image domain, and the data acquisition of diseased images is extremely expensive because of the  privacy issues of patients. Furthermore, sometimes the incidence of some diseases is extremely rare.
In contrast, it is relatively easier to collect the normal (healthy) data.
Human can distinguish those images with diseases from normal healthy data, and it is important for an intelligent system to mimic the behavior of the human for detecting those images with diseases by only leveraging the normal training data, and such task is defined as anomaly detection \cite{schlegl2017unsupervised} in medical image analysis domain.

\begin{figure*}[ttt]
	\centering
	\includegraphics[width=\textwidth]{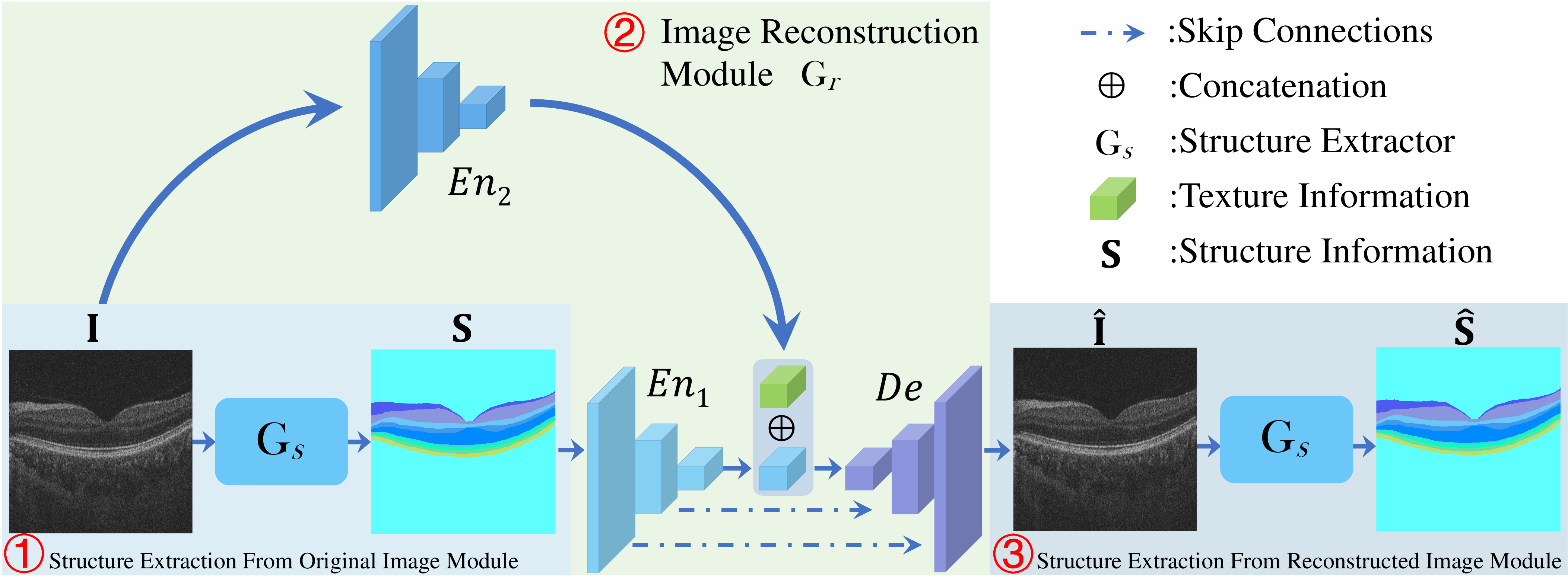}
	\vspace{-0.05in}
	\caption{The pipeline of our P-Net, which consists of three modules. Firstly, the structure extraction network $\mathbf{G}_s$ is trained for extracting structure $\mathbf{S}$ from the original image $\mathbf{I}$, and then the extracted $\mathbf{S}$ and feature encoded from $\mathbf{I}$ are fused for reconstruction. Finally, we further utilize the reconstructed image $\mathbf{\hat{I}}$ to extract the $\mathbf{\hat{S}}$ and measure the difference between $\mathbf{S}$ and $\mathbf{\hat{S}}$. Our P-Net encodes the relation between image texture and structure by enforcing the consistency of the image and structure between original and reconstructed ones.
	}
	\label{fig:method_overall}
	\vspace{-0.10in}
\end{figure*}

Typical anomaly detection methods are usually based on image reconstruction \cite{baur2018deep}\cite{chen2018unsupervised}\cite{schlegl2019f}\cite{zhou2017anomaly}\cite{zimmerer2018context}\cite{zhou2020sparse}
which means given an image, an encoder maps the image to feature space and a decoder reconstructs the image based on the feature. By minimizing the reconstruction error between the input image and the reconstructed image on the normal training data, the encoder and decoder are trained for image reconstruction.
In the testing phase, an image can be classified as normal or abnormal by measuring the reconstruction error \cite{zhou2017anomaly}\cite{zimmerer2018context}. To guarantee the fidelity of the reconstructed image on the normal training data, generative adversarial networks (GAN) \cite{goodfellow2014generative} based solutions have been introduced \cite{baur2018deep}\cite{chen2018unsupervised},
which guide the generator to synthesize more realistic images with a discriminator.
Further, GANomaly \cite{akcay2018ganomaly} and f-AnoGAN \cite{schlegl2019f} are proposed, which append an additional encoder  to the generator to further encode the reconstructed image. Then the reconstruction errors corresponding both images and features to measure the anomaly.
However, all these existing methods directly feed the image into the CNNs for anomaly detection without leveraging any prior information. When doctors make diagnosis, besides the textures of the organ in the image, the structures (here \textbf{we treat the semantically meaningful edges in an image as the structure}, e.g., vessel topological structure in fundus images, the anatomic layer structure in OCT images, \emph{etc.}) also help them to make the decision \cite{puliafito1995imaging}\cite{zinreich1988fungal}\cite{hartnett1996deep}.
As shown in Fig. \ref{fig:intro_structure}, for eye images with diseases, the normal structures are destroyed.
For normal (healthy) images, the structure can be extracted, and the extracted structure also provides a cue about texture distribution. Since both texture and structure helps anomaly detection, then a question is naturally raised: \textit{How to encode the structure-texture relation with CNNs for anomaly detection?} Towards this end, we propose to leverage the dependencies between structure and image texture for image and structure reconstruction circularly, and use use the reconstruction error for both structure and image as normality measurement.

Specifically, we first propose to extract the structure from the original image, then we map the structure to the reconstructed image. However, the mapping from the structure to the reconstructed image is ill-posed.
Thus we propose to fuse the last layer image feature with structure feature to reconstruct the image.
We further use the reconstructed image to extract the structure, which also serves as a regularizer and helps improve the image reconstruction in previous stage. Meanwhile, the structure difference between the structure extracted from original image and that from the reconstructed image also helps us to measure the anomaly score. As shown in Fig. \ref{fig:method_overall}, since the whole network architecture is like a ``P'', we term it as P-Net.

In the training phase, since the structure of retinal image are usually not given for anomaly detection, we propose to use the vessel segmentation datasets and OCT layer segmentation datasets to train the structure extraction module of our network with a domain adaption method \cite{chai2020perceptual}. By minimizing the error between the input image and its reconstructed version (referred to as contents error), and the error between the structure extracted from original image and that extracted from the reconstructed image (referred to as structure error), our P-Net can be trained. In the inference stage, by measuring the contents error and structure error, each image can be classified as normal/abnormal accordingly. It is worth noting that our retinal image anomaly detection approach is a general framework, it can be readily applied to anomaly detection for general object images and novel class discovery for retinal images where testing data contains data falling out of the distribution of the training data \footnote{These tasks are also termed as general anomaly detection in computer vision}. For example, training data contains some given types of diseases while testing data contains a new type of disease. The reason for the success of our P-Net in these cases is that our network can capture the consistency between the structure and image contents, and the structure and image contents relation is different from the training ones for those new diseases.

The main \textbf{contributions} of this work are summarized as follows: 
\textbf{i)} we propose to utilize the structure information for anomaly detection in the retinal image. Our solution agrees with the cognition of clinicians that the normal retinal images usually have regular structures, and the irregular structure hints the incidence of some diseases.  To the best of our knowledge, this is the first work that infuses structure information into CNNs for anomaly detection;
\textbf{ii)} we propose a novel P-Net that encodes the relation between structure and textures for anomaly detection by using the cycle reconstruction between the image contents and structure. In the inference stage, both image reconstruction error and structure difference are utilized for anomaly score measurement;
\textbf{iii)} since the structures are not given on almost all anomaly detection datasets for retinal images, we employ a domain adaptation method to extract structure by leveraging other datasets annotated with structure; 
\textbf{iv)} extensive experiments validate the effectiveness of our method for anomaly detection in both fundus modality and OCT modality for retinal images. Further, our method can be well generalized to novel class discovery for retinal images and anomaly detection for general object images.

\section{Related Work}
\vspace{-0.05in}
\subsection{Anomaly Detection}
Anomaly detection is a vital field in the machine learning. An intuitive assumption is that the anomalies are out of the distribution of normal samples. 
Based on this hypothesis, it is natural to learn a discriminative hyperplane to separate the abnormal samples from the normal ones. 
One-class support vector machine (OCSVM) \cite{scholkopf2000support} was one of the classical methods, and its derived deep one-class SVDD \cite{ruff2018deep} constrained the normal samples in a hypersphere so that the potential anomalies are the outliers being far away from the center of the hypersphere. 
Besides, Gaussian Mixture Models (GMM) tends to model the distribution of normal samples, and the outliers out of the distribution might result in a high probability of being abnormal.
 
Schlegl \etal \cite{schlegl2017unsupervised} initially introduced Generative Adversarial Networks (GANs) \cite{goodfellow2014generative} for anomaly detection that termed AnoGAN.
The AnoGAN generates images from a Gaussian latent space, and samples are recognized as anomalies when the corresponding latent code is out of the distribution. 
Similar to AnoGAN, GANomaly \cite{akcay2018ganomaly} also involved representation learning in latent space. Compared with AnoGAN, GANomaly does not seek the latent code in the manifold by gradient descent in the test phase.
David \etal \cite{zimmerer2018context} proposed the context-encoding Variational Auto-Encoder in brain MRI images, which combines reconstruction with density-based anomaly scoring. 
Schlegl \etal \cite{schlegl2019f} proposed to utilize a generator \cite{goodfellow2014generative} to map latent space to normal retinal OCT image, and use an encoder to learn the mapping from retinal OCT image to latent space.
Pramuditha \etal \cite{perera2019ocgan} proposed OCGAN to make all the samples in a closed latent space. It reconstructs all samples to the normal ones. 
Also, the memory-augmented network such as \cite{gong2019memorizing} provided a fascinating idea to map the latent code of each sample to the nearest item in a learned dictionary with only normal patterns. 

As discussed before, for normal healthy images, the structure and image texture are closely related. However, these existing methods fail to encode the structure-texture relation. 

\subsection{Structure-Texture Relation Encoding Networks}
The texture and structure in an image are complementary to each other \cite{aujol2006structure}, and image structure has been successful used for image inpainting \cite{Ren_2019_ICCV}\cite{Nazeri_2019_ICCV_Workshops}. 
Nazeri \etal \cite{Nazeri_2019_ICCV_Workshops} proposed a two-stage network, which took the edge information as the structure. The model  \cite{Nazeri_2019_ICCV_Workshops} first predicts the full edge map of incomplete image by the edge generator. Then the predicted edge map and incomplete image are passed to an image completion network to compute the full image. Since the distribution of edge map is significantly  different from the distribution of the color image, Ren \etal  \cite{Ren_2019_ICCV} proposed to employ edge-preserved smooth images to represents the structure of the color images. The network proposed in \cite{Ren_2019_ICCV} consists of a structure reconstructor to predict the image structure and a texture generator to complete the image texture. The relation of structure-texture can be encoded in the `image-structure-image' pipeline, and this motivates us to infuse the normal structure into deep neural networks for anomaly detection. 
In our work, we further encode the relation between normal image and structure by enforcing the consistency of normal image, and the consistency between the structure extracted from normal image and that extracted from the reconstructed image.


\section{Method}
\label{method}

For healthy populations, the distribution of vasculature and histology of the retinal layers is regular. On the contrary, for subjects with diseases, the lesion of diseases will destroy the regularity of vasculature and histology.
For example, 
the blood vessel and histology layer in retina will be destroyed by diabetic retinopathy (DR).
The layer-wise structure in OCT will also be destroyed by various lesions such as pigment epithelium detachment (PED), subretinal fluid (SRF) \cite{hu2019automated}, \textit{etc}. 

Based on these clinical observations, we define the retinal blood vessels in fundus images and the retinal layers in OCT as structure. Besides the anomalies in texture, the anomalies of structure would also help ophthalmologists and clinicians to make the diagnosis decision \cite{puliafito1995imaging}\cite{hartnett1996deep}.
Motivated by the functionality of structure in retinal disease diagnosis, we propose to leverage the structure as an additional cue for anomaly detection. Further, for healthy images, structure extracted from the image provides a cue about texture distribution. By leveraging the relation between structure and texture in retinal images, we propose a P-Net for anomaly detection, and P-Net encodes the dependencies between structure and relation.


Specifically, our network architecture consists of three modules: 1) structure extraction from original image module, denoted as $\mathbf{G}_{s}$, which extracts structure $\mathbf{S}$ from original image $\mathbf{I}$; 2) image reconstruction module, denoted as $\mathbf{G}_{r}$, which leverages the last layer image encoder feature and structure to reconstruct the input image. We denote the reconstructed image as $\mathbf{\hat{I}}$. By minimizing the difference between $\mathbf{I}$ and $\mathbf{\hat{I}}$, the relation between texture and structure is encoded into the network. Thus we use image reconstruction error ($\|\mathbf{I}-\mathbf{\hat{I}}\|_1$) as a normality measurement;
3) structure extraction from reconstructed image module, which further extracts structure from the reconstructed image $\mathbf{\hat{I}}$. We denote the structure extracted from $\mathbf{\hat{I}}$ as $\mathbf{\hat{S}}$. By minimizing the difference between $\mathbf{S}$ and $\mathbf{\hat{S}}$, this module enforces the original image to be correctly reconstructed by $\mathbf{G}_{r}$. Further, the structure difference $\|\mathbf{S}-\mathbf{\hat{S}}\|_1$) can also be used to measure the normality of the image. The network architecture of P-Net is shown in Fig. \ref{fig:method_overall}. The detailed architecture of each module can be found in the supplementary (Section S1).

\subsection{Structure Extraction From Original Image Module}


\begin{figure}[htb]
	\centering
	\includegraphics[width=\textwidth]{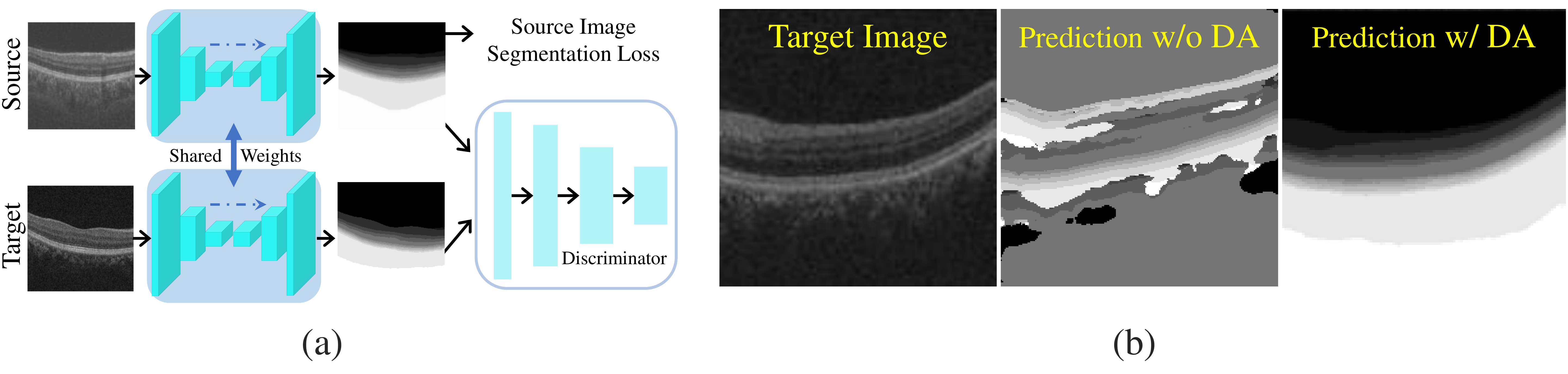} 
	\caption{(a) Structure extraction network with domain adaptation (DA). (b) The qualitative results of DA for OCT images. The structure of target image cannot be extracted well without DA.}
	\label{fig:method_sem}
\end{figure}


The datasets used in previous retinal image anomaly detection work \cite{schlegl2017unsupervised}\cite{schlegl2019f} are not publicly available, therefore we propose to use the Retinal Edema Segmentation Challenge Dataset (RESC, a OCT image dataset) \cite{hu2019automated}, and a fundus multi-disease diagnosis dataset (iSee dataset \cite{yan2019oversampling}) collected in a local hospital for performance evaluation. However, the structure in both datasets are not provided. Manually annotating the structure, including vessels in fundus and layer segmentation in OCT, is extremely time consuming.
Fortunately, there are many publicly available datasets for vessel segmentation in fundus images and layer segmentation in OCT images \cite{hoover2000locating}\cite{staal2004ridge}.
To get structure without tedious manual annotation, we utilize existing datasets to train a network for structure extraction.
However, retinal images in different datasets are captured by various devices, consequently, different datasets have different noises and data distribution.
To tackle this problem, we leverage AdaSeg \cite{tsai2018learning}, a domain adaptation based image segmentation method to learn the structure extractor $\mathbf{G}_{s}$. Specifically, we map images in different datasets but with the same modality to their corresponding structures with a U-Net \cite{ronneberger2015u}, and add a discriminator to make the segmentation results from source and target datasets indistinguishable. The network architecture is shown in Fig. \ref{fig:method_sem}(a). For RESC, we use the Topcon dataset \cite{cheng2016speckle} as the source; while for iSee, we use the DRIVE dataset \cite{staal2004ridge} as the source. The training loss in this module is as follows:

\begin{align}
\mathcal{L}_{\text{seg}}(I_{\text{src}}) &= -\sum S_{\text{src}} \log(\mathbf{G}_{s}(I_{\text{src}}))	\\
\mathcal{L}_{\text{seg}}(I_{\text{tar}}) &= \mathbb{E}[\log(1-D(\mathbf{G}_{s}(I_{\text{tar}}))] + \mathbb{E}[\log D(\mathbf{G}_{s}(I_{\text{src}}))]
\end{align}
where $I_{\text{src}}$ and $S_{\text{src}}$ denote the source image and its ground truth, respectively. $I_{\text{tar}}$ denotes the target image, and $D$ denotes the discriminator.
Once the structure extraction module is trained, we fix the module to simplify the optimization of the other modules in our P-Net.


\subsection{Image Reconstruction Module}
Since the structure is represented by vessels in fundus or layer section in OCT, and the ambiguity exists for the direct mapping from structure to original image \cite{Ren_2019_ICCV}\cite{isola2017image}, we propose to combine structure information and image texture information to reconstruct the original image. We define the texture as complementary information of the structure, and the texture provides the details over the local regions.

Specifically, we encode the original image and its structure with $\text{En}_1$ and $\text{En}_2$, respectively. Then we concatenate the two features and feed them into a decoder ($\text{De}$) to reconstruct the original image. Skip connections are introduced between the structure encoder and decoder for features at the same level, which avoid the information loss caused by downsampling pooling in structures, while there is no skip connection between the image encoder and decoder. The reason is that if we introduce the skip connection between them, then it is possible that we learn an identity mapping between the image and reconstructed image, which leads that there is no information flowed from structure to the original image, which is not desirable because identity mapping also makes abnormal images well reconstructed in the testing phase \cite{chen2018deep}, where anomaly detection is impossible. It is expected that only information related to texture is encoded in the original encoder and passed to decoder to help the image reconstruction, therefore probably the last layer feature in the image encoder is enough for this purpose.

Following \cite{akcay2018ganomaly}\cite{isola2017image}, we use $L_1$ norm to measure the difference between the reconstructed image and the original image. 

\begin{equation}
\mathcal{L}_{\text{rec}}(\mathbf{I}) = \| \mathbf{I} - \mathbf{\hat{I}}\|_1
\end{equation}

To improve the quality of the reconstructed image, we apply PatchGAN \cite{isola2017image} to penalize the reconstruction error for the reconstructed image $\mathbf{\hat{I}}$. Formally, let $\mathbf{D}$ be the discriminator, the adversarial loss $\mathcal{L}_{\text{adv}}$ for training reconstruction network is shown as follows:

\begin{equation}
\mathcal{L}_{\text{adv}}(\mathbf{I}) = \mathbb{E}[\log (1-\mathbf{D}(\mathbf{G}_{r}(\mathbf{I},\mathbf{S})))] + 
\mathbb{E}[\log \mathbf{D}(\mathbf{I})]
\end{equation}

\subsection{Structure Extraction From Reconstructed Image Module}
We further append the structure extractor $\mathbf{G}_{r}$ to the reconstructed image. There are two purposes: 1) by enforcing the structure extracted from original image
and that from reconstructed image
to be the same, the original image can better reconstructed. In this sense, image reconstruction from reconstructed image module behaves like a regularizer; 2) some lesions are more discriminative in structure, then we extract structure from original image and reconstructed image, respectively, and use their difference for normality measurement. The loss function in this module is defined as follows:
\begin{equation}
\mathcal{L}_{\text{str}}(\mathbf{I}) = \| \mathbf{S} - \mathbf{\hat{S}} \|_1
\end{equation}
 
\subsection{Objective Function}
We fix the structure extractor $\mathbf{G}_{s}$ in the training of image reconstruction module $\mathbf{G}_{r}$.
Therefore, we arrive at the the objective function of our P-Net:
\begin{equation}
\mathcal{L} = \lambda_1 \mathcal{L}_{\text{adv}} + \lambda_2 \mathcal{L}_{\text{rec}} + \lambda_s \mathcal{L}_{\text{str}}
\end{equation}
where $\lambda_1, \lambda_2, \lambda_s$ are the hyper-parameters. Empirically, we set $\lambda_1=0.1, \lambda_2=1, \lambda_s=0.5$ on all datasets in our experiments.

\subsection{Anomaly Detection for Testing Data}

%
We combine image reconstruction error with structure difference for anomaly score ($\mathcal{A}(\mathbf{I})$) measurement:
\begin{equation}
\mathcal{A}(\mathbf{I}) = (1 - \lambda_f) \| \mathbf{I} - \mathbf{\hat{I}} \|_1 + \lambda_f \|\mathbf{S} - \mathbf{\hat{S}} \|_1
\label{equa:anomaly_fusion}
\end{equation}
where $\lambda_{f}$ is a weight used to balance the image difference and structure difference.
A higher anomaly score indicates that the image is more likely to be abnormal. 

\section{Experiments}
\subsection{Implementation}
To train the network, the input image size is 224 $\times$ 224, and the batch size is 8. The optimizer for the generator and the discriminator are both Adam, and the learning rate is 0.001. We train our model for 800 epochs.
We implement our method with the PyTorch on a NVIDIA TITAN V GPU. The codes are released in
\url{https://github.com/ClancyZhou/P_Net_Anomaly_Detection}.

\subsection{Evaluation Metric}
Following previous work \cite{zhou2020sparse}\cite{luo2017revisit}\cite{luo2019video}, we calculate the Area Under Receiver Operation Characteristic (AUC) by gradually changing the threshold of $\mathcal{A}(\mathbf{I})$ for normal/abnormal classification. A higher AUC indicates that the performance of the method is better.


\subsection{Anomaly Detection in Retinal Images}
\noindent
\textbf{A. Datasets.}
Since the datasets used in previous retinal image anomaly detection work \cite{schlegl2019f}\cite{schlegl2017unsupervised} are not released, we evaluate our proposed method with a publicly available dataset \cite{hu2019automated} and a local hospital dataset \cite{yan2019oversampling}.

\noindent\textbf{Retinal Edema Segmentation Challenge Dataset (RESC)}\cite{hu2019automated}.
Retinal edema is a retinal disease, which causes blurry vision and affects the patient's life quality. Optical coherence tomography (OCT) images can be used to assist clinicians in diagnosing retinal edema. Thus the RESC dataset is proposed for OCT based retinal edema segmentation. As discussed previously, retinal edema damages the normal layer structure in OCT, thus we leverage this dataset for performance evaluation. This dataset contains the standard training/validating split. We use the normal images in the original training set as our training images to train the model, and use all testing images for performance evaluation.


\noindent\textbf{Fundus Multi-disease Diagnosis Dataset (iSee)}\cite{yan2019oversampling}.
Previous retinal fundus datasets usually only contain one or two types of disease \cite{porwal2019idrid}\cite{orlando2020refuge}, but in clinical diagnosis, many eye diseases can be observed in the fundus image.
Thus we collect a dataset from a local hospital, which comprises of 10000 fundus images. Eye diseases in this dataset include age-related macular degeneration (AMD), pathological myopia (PM), glaucoma, diabetic retinopathy (DR), and some other types of eye diseases.
To validate the effectiveness of P-Net for different retinal diseases, we use 4000 normal images as the training set, and we use the remaining 3000 normal images, 700 images with AMD, 800 images with PM, 420 images with glaucoma, 480 images with DR, and 600 images with other types of eye diseases, as our testing set.

\noindent
\textbf{B. Performance Evaluation.}

\begin{table}[htb]
	\scriptsize
	\centering
	\caption{Performance comparison on different datasets.}
	\begin{tabular}{p{1.2in}<{\centering}|p{0.8in}<{\centering}|p{0.8in}<{\centering}}
		\hline
		Method & RESC (OCT) & iSee (fundus) \\ \hline
		Deep SVDD \cite{ruff2018deep} & 0.7440 & 0.6059 \\
		Auto-Encoder \cite{zhou2017anomaly} & 0.8207 & 0.6127 \\
		AnoGAN \cite{schlegl2017unsupervised}  & 0.8481  & 0.6325 \\
		VAE-GAN \cite{baur2018deep} & 0.9064 & 0.6969 \\
		Pix2Pix \cite{isola2017image} & 0.7934 & 0.6722 \\
		GANomaly \cite{akcay2018ganomaly} & 0.9196  & 0.7015 \\
		Cycle-GAN \cite{zhu2017unpaired} & 0.8739 & 0.6699 \\ \hline
		Our Method & \textbf{0.9288}  & \textbf{0.7245}  \\ \hline
	\end{tabular}
	\label{table:retinal_com}
	\vspace{-0.15in}
\end{table}

\textbf{Baselines.} We compare our method with AnoGAN \cite{chen2018unsupervised} proposed for retinal OCT images, VAE-GAN \cite{baur2018deep} proposed for Brain MRI images, GANomaly \cite{akcay2018ganomaly} for X-ray security images, and Auto-Encoder based anomaly detection \cite{zhou2017anomaly}. As our work consists of the translation between image to the structure. Therefore we also compare P-Net with image-to-image translation networks, including Pix2Pix \cite{isola2017image} and Cycle-GAN \cite{zhu2017unpaired}. For Pix2Pix \cite{isola2017image} and Cycle-GAN \cite{zhu2017unpaired}, we use the original image and structures extracted with domain adaptation method to train the network, and use the same measurement as ours for anomaly detection.

As shown in Table \ref{table:retinal_com}, our method outperforms all baseline methods on both datasets, which verifies the effectiveness of our method for retinal images with different modalities.

\begin{table}[htb]
	\vspace{-0.15in}
	\centering
	\scriptsize
	\caption{The results of sub-class on iSee dataset. }
	\begin{tabular}{p{1.0in}<{\centering}|p{0.45in}<{\centering}p{0.45in}<{\centering}p{0.45in}<{\centering}p{0.45in}<{\centering}p{0.45in}<{\centering}}
		\hline
		Method & AMD & PM & Glaucoma & DR & Other \\  \hline
		Auto-Encoder \cite{zhou2017anomaly} & 0.5463 & 0.7479 & 0.5604 & 0.6002 & 0.5479 \\
		AnoGAN \cite{schlegl2017unsupervised}  & 0.5630 & 0.7499 & 0.5731 & 0.5704 & 0.6412 \\ 
		VAE-GAN \cite{baur2018deep} & 0.5593 & 0.8412 & \textbf{0.6149} & 0.6590 & 0.7961  \\
		GANomaly \cite{akcay2018ganomaly} & \textbf{0.5713} & 0.8336 & 0.6056 & 0.6627 & 0.8013  \\ \hline
		Our Method & 0.5688 & \textbf{0.8726} & 0.6103 & \textbf{0.6830} & \textbf{0.8069} \\ \hline
	\end{tabular}
	\label{tab:sub_cls}
	\vspace{-0.15in}
\end{table}

We further report the AUC of our method for five sub-classes in the iSee dataset, i.e., AMD, PM, glaucoma, DR, and other disease classes, and show the results in Table \ref{tab:sub_cls}, As the lesions of PM and DR are related to blood vessel structure,
and our method encodes the relation between vessels and texture, our method performs well for these diseases. While the lesions of AMD and glaucoma are associated with the macular area and the optic disc, respectively, and the structures we used cannot cover these areas in our current implementation, therefore our solution doesn't perform very well for these diseases.

\begin{figure*}[htb]
	\centering
	\includegraphics[width=0.90\textwidth]{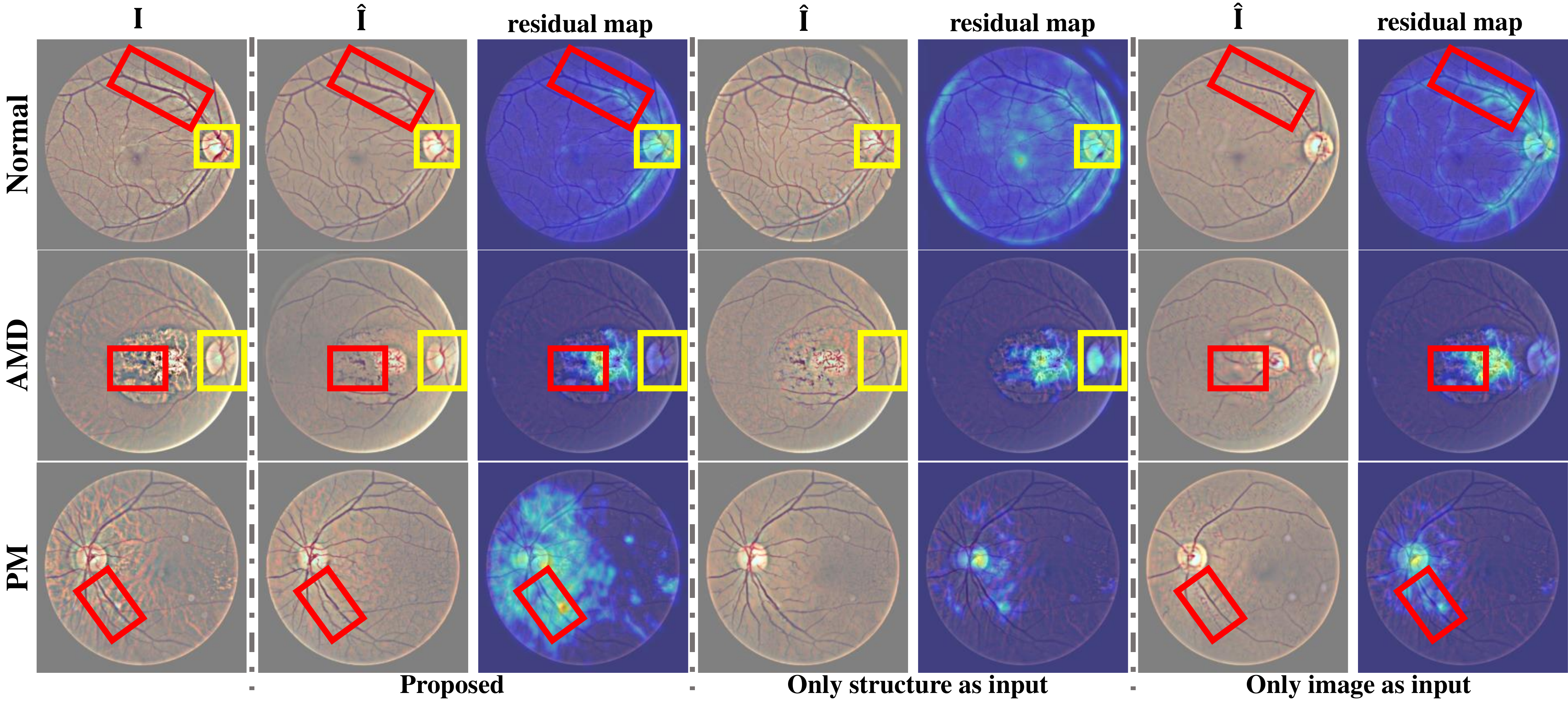}
	\vspace{-0.05in}
	\caption{The qualitative results of different input of $\mathbf{G}_r$ show that \textbf{both image and structure are necessary for anomaly detection}. 
		On the one hand, when the input is only structure, we can observe that optical disc (yellow box) cannot be reconstructed precisely. The reason is the lack of texture information of optical disc region. 
		On the other hand, when the input is only image, the lack of structure (blood vessel) information results in the vessel information loss (red box of normal and PM), and the lesion region is reconstructed incorrectly as the vessel (red box of AMD).
	}
	\label{fig:exp_ablation}
	\vspace{-0.05in}
\end{figure*}


\noindent
\textbf{C. Ablation Study.}


\begin{table}[htb]
	\centering
	\scriptsize
	\caption{Ablation study in different datasets. DA and $\mathcal{L}_{\text{str}}$ denote domain adaptation, and structure consistency loss respectively.}
	\begin{tabular}{p{0.4in}<{\centering}|p{0.4in}<{\centering}|p{0.4in}<{\centering}p{0.4in}<{\centering}|p{0.4in}<{\centering}|p{0.45in}<{\centering}p{0.45in}<{\centering}}
		\hline	
		 & \multicolumn{1}{c|}{\multirow{2}{*}{DA}} & \multicolumn{2}{c|}{Input of $\mathbf{G}_{r}$} & \multicolumn{1}{c|}{\multirow{2}{*}{$\mathcal{L}_{\text{str}}$ }} & \multicolumn{2}{c}{AUC} \\ \cline{3-4} \cline{6-7}
		Index & \multicolumn{1}{c|}{} & Image & \multicolumn{1}{c|}{Structure} & \multicolumn{1}{c|}{} & RESC & iSee \\ \hline
		
		1 &	  		 	& \checkmark & \checkmark &  & 0.8152 & 0.5914 \\ \hline
		2 & \checkmark 	& \checkmark &  &  & 0.8219 & 0.6487 \\ \hline
		3 &	\checkmark 	& & \checkmark &  & 0.8277 & 0.6914 \\ \hline
		4 &	\checkmark 	& \checkmark & \checkmark &   & 0.8518 & 0.7196 \\ \hline
		5 &	\checkmark 	& \checkmark &  & \checkmark & 0.8835 & 0.6574 \\ \hline
		6 &	\checkmark 	& & \checkmark & \checkmark & 0.8821 & 0.6993 \\ \hline
		Ours& \checkmark & \checkmark &  \checkmark & \checkmark & \textbf{0.9288} & \textbf{0.7245} \\ \hline
	\end{tabular}
	\label{tab:ablation}	
	
\end{table}

\textbf{Domain Adaptation (DA).}
As shown in Fig. 
\ref{fig:method_sem}(b)
since there is domain discrepancy between source images and target images, if we train a segmentation model without domain adaptation, the quality of structure is not good enough for image reconstruction. The quantitative results are listed in Table \ref{tab:ablation} (row 1 vs. row 4). We can see that our method benefits from DA on both datasets.

\textbf{The Input of $\mathbf{G}_{r}$}. Our P-Net takes both the structure map and the original image as input for reconstruction.
To investigate the effectiveness of this design, we conduct qualitative and quantitative experiments. The results are shown in Fig. \ref{fig:exp_ablation} and Table \ref{tab:ablation} (row 2, 3, and 4), respectively. As shown in Table \ref{tab:ablation}, our P-Net solution is better than single input based image reconstruction strategy.
Further, from Fig. \ref{fig:exp_ablation}, we can observe that: i) if $\mathbf{G}_{r}$ only takes the structure as input, the image texture such as the optic disc area will be poorly reconstructed; ii) if $\mathbf{G}_{r}$ only takes the image as input, the blood vessel and macular area will be poorly reconstructed, for example, the macular area is reconstructed as the blood vessel, which is obviously incorrect.


\textbf{Structure Consistency Loss.}
The $\mathcal{L}_{\text{str}}$ constrains the consistency between  $\mathbf{\hat{S}}$ and $\mathbf{S}$, which behaves like a regularizer to enforce the consistency between  $\mathbf{\hat{I}}$ and $\mathbf{I}$.
The results in Table \ref{tab:ablation} (row 2 vs. row 5, row 3 vs. row 6, and row 4 vs. row 7) validate the effectiveness of the $\mathcal{L}_{\text{str}}$.

\noindent
\textbf{D. Evaluation of $\lambda_f$.}

In the testing phase, we use Equation (\ref{equa:anomaly_fusion}) to measure the anomaly score. $\lambda_f = 0$ denotes that only image difference $\| \mathbf{I} - \mathbf{\hat{I}} \|_1$ is used for anomaly detection, and $\lambda_f = 1$ means only structure difference $\| \mathbf{S} - \mathbf{\hat{S}} \|_1$ is used for anomaly detection. We vary $\lambda_f$ and show the results in Table \ref{table:lambda_f}. We can see that the performance of image difference only based anomaly detection is worse than structure difference only based method. The possible reason is that the structure is more evident for anomaly detection, which agrees with practice of clinicians. If we combine both, then it leads to a better performance. Further, the model is robust to different $\lambda_f$'s. When $\lambda_f = 0.8$, our proposed method achieves the best performance, thus we set $\lambda_f = 0.8$ in all the experiments.

\begin{table}[htb]
	\centering
	\scriptsize
	\caption{Results of different $\lambda_f$ on RESC dataset.}
	\begin{tabular}{cccccccc}
		\hline
		$\lambda_f$ & 0.0 & 0.2 & 0.4 & 0.5 & 0.6 & \textbf{0.8} & 1.0 \\ \hline
		AUC & 0.8481 & 0.9010 & 0.9226 & 0.9232 & 0.9234 & \textbf{0.9288} & 0.9084 \\\hline
	\end{tabular}
	\label{table:lambda_f}
\end{table}

\noindent
\textbf{E. The Number of Cycles in P-Net.}

\begin{figure}[htb]
	\centering
	\includegraphics[width=0.90\textwidth]{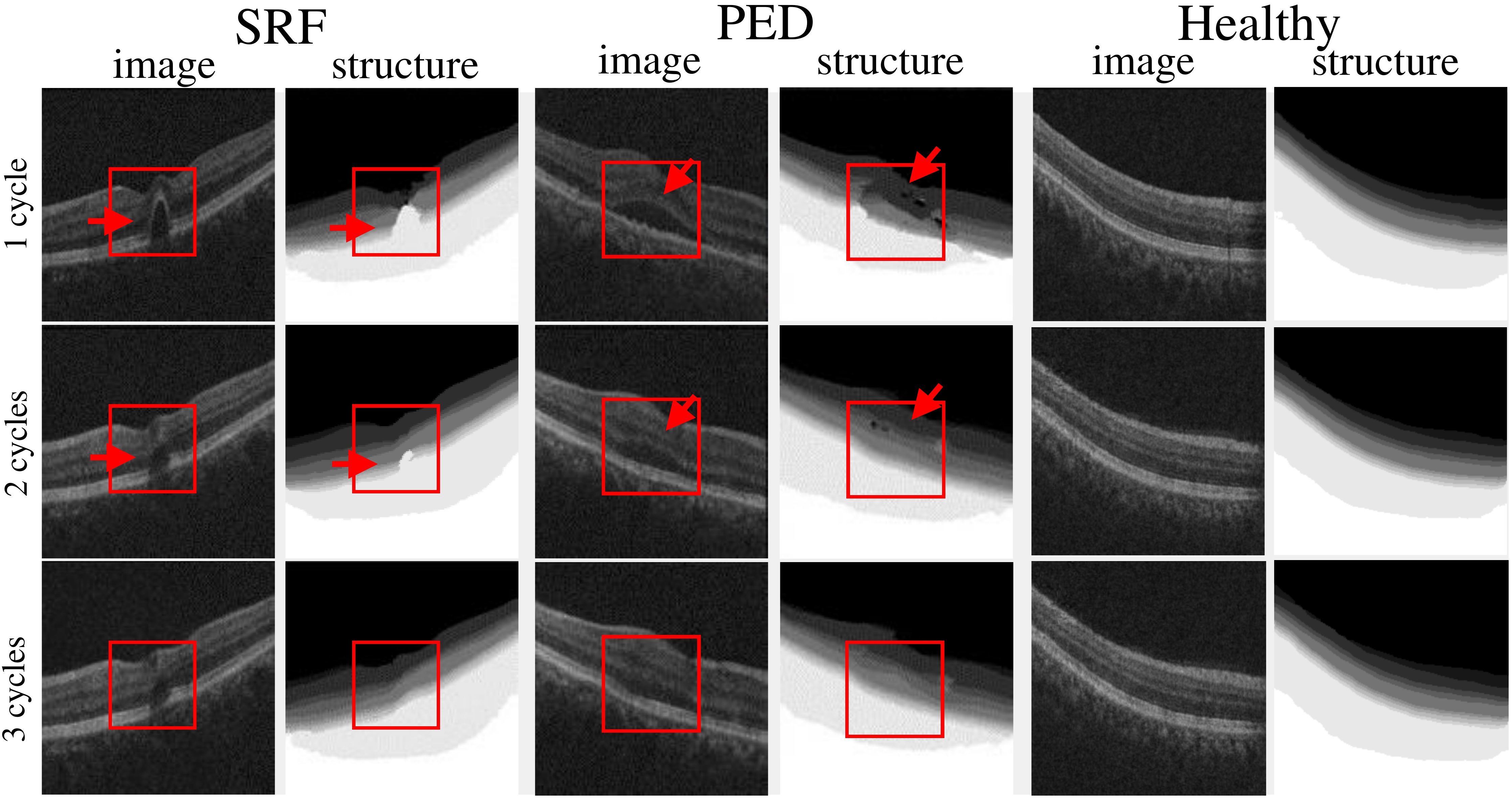}
	\vspace{-0.05in}
	\caption{The qualitative results with different number of cycles. It can be observed that the texture and the layer-wise structure between original and reconstructed ones in the healthy sample are consistent, while the consistency in abnormal samples is broken.
	}
	\label{fig:exp_cycle}
	\vspace{-0.10in}
\end{figure}

The lesion areas cannot be well reconstructed, and we found that these reconstructed lesion areas are very similar to normal areas. 
Thus, the results of appending our framework several times are different from that with only one cycle. We show the distribution of reconstruction errors for both normal images and abnormal images with multiple cycles reconstruction.
It shows that more cycles in testing phase improve the performance, while more cycles in training phase reduce the performance.
The poor performance of more cycles in training phase is probably because more loss terms make the optimization more difficult.

In Fig. \ref{fig:exp_cycle}, we further show the qualitative effect of more cycles in testing phase, where the images correspond to pigment epithelium detachment (PED), subretinal fluid (SRF), and healthy image, respectively.
In the multiple cycles in testing phase, the abnormal lesion becomes more and more similar to normal patterns, which is a little like ``anomalies repairing''. Such phenomenon is more obvious in the structure map.
For the healthy image, both the image and structure map remain the same even after multiple cycles. Thus more cycles would enlarge the reconstruction error for abnormal images, and retain the same reconstruction error for normal ones, which explains the phenomenon that more cycles in testing phase improve the anomaly detection.

\begin{table}[htb]
	\vspace{0.05in}
	\centering
	\scriptsize
	\caption{The results of different number of cycle for training and testing on RESC dataset.}
	\begin{tabular}{c|ccccc}
		\hline
		Cycle number in test &   1 &  2 &   3 &   4 &  5\\ \hline
		1 cycle in train & 0.9288 & 0.9304 & 0.9361 & 0.9380 & 0.9374\\
		2 cycles in train & 0.8935 & 0.8962 & 0.9022 & 0.8973 & 0.9015 \\
		\hline
	\end{tabular}
	\label{tab:cycle_oct}
	\vspace{-0.15in}
\end{table}

\subsection{Anomaly Detection in Real World Images}
\renewcommand{\arraystretch}{1}
We also apply our method on the MVTec AD dataset \cite{bergmann2019mvtec}, which is a very challenging and comprehensive anomaly detection dataset for general object and texture images. This dataset contains 5 texture categories and 10 object categories. Since the structure is not annotated on these real-world images, we simply take the edges detected by Canny edge detection as the structure. We compare our method with Auto-Encoder with L2 loss or SSIM loss \cite{bergmann2019mvtec}, CNN Feature Dictionary (CFD) \cite{napoletano2018anomaly}, Texture Inspection (TI) \cite{bottger2016real}, AnoGAN \cite{schlegl2017unsupervised}, Deep SVDD \cite{ruff2018deep}, Cycle-GAN \cite{zhu2017unpaired}, VAE-GAN \cite{baur2018deep} and GANomaly \cite{akcay2018ganomaly}.
The quantitative results are shown in Table \ref{tab:mvtec_results}, and qualitative results are provided in the supplementary (Fig. S1).

We utilize AUC and region overlap as evaluation metrics to evaluate the performance of our model on MVTec AD dataset. Following \cite{bergmann2019mvtec}, we define a minimum defect area for normal class data. Then we segment the difference map of normal class samples with increased threshold. This process is not stopped until the area of anomaly region is just below the defect area we defined and this threshold is utilized for segmentation anomaly region in testing phase.

We can see our method achieves the best performance in terms of the average AUC and average anomaly region overlap on all categories. Further, our method is effective for object images and less effective for some type of texture images. The possible reason is that we use the edge as structure. For object images, such edges usually correspond to shapes, which is closely related to the image contents. Thus the mapping between image and structure is relatively easy, which consequently helps the anomaly detection. For abnormal object images, usually some parts are broken or missing, which would leads to a large reconstruction error.
However, since there are too many edges in texture images and the edges are very noisy, the texture image is hard to reconstruct, consequently reduces the performance of anomaly detection.

The experimental results of novel class discovery are provided in the supplementary (Section. S3).

\begin{table}[h]
	\scriptsize
	\caption{For each category, the top row is the anomaly region \textbf{overlap} which is the same as the evaluation metric in \cite{bergmann2019mvtec} and the bottom row is \textbf{AUC}. The 5 categories at the top of the table are textures 
		image and the other 10 categories at the bottom of the table is objects 
		image. The results of AE (SSIM), AE (L2), AnoGAN \cite{schlegl2017unsupervised}, CFD \cite{napoletano2018anomaly}, and TI \cite{bottger2016real} are adopted from MVTec AD \cite{bergmann2019mvtec} dataset directly.}
	\centering
	\begin{tabular}{cp{0.36in}<{\centering}p{0.36in}<{\centering}p{0.36in}<{\centering}p{0.36in}<{\centering}p{0.36in}<{\centering}p{0.36in}<{\centering}p{0.36in}<{\centering}p{0.36in}<{\centering}p{0.36in}<{\centering}p{0.36in}<{\centering}}
		\hline
		Categories & \begin{tabular}[c]{@{}c@{}}AE\\ (SSIM)\end{tabular} & \begin{tabular}[c]{@{}c@{}}AE\\ (L2)\end{tabular} & \begin{tabular}[c]{@{}c@{}}Ano\\ GAN\end{tabular} & CFD & \begin{tabular}[c]{@{}c@{}}Deep\\ SVDD\end{tabular} 
		& \begin{tabular}[c]{@{}c@{}}Cycle\\ GAN\end{tabular}
		& \begin{tabular}[c]{@{}c@{}}VAE-\\ GAN\end{tabular}
		& \begin{tabular}[c]{@{}c@{}}GAN\\omaly\end{tabular}
		& TI & \begin{tabular}[c]{@{}c@{}}Our\\ Method\end{tabular} \\ \hline 
		& \textbf{0.69} & 0.38 & 0.34 & 0.20 & - & 0.04 & 0.01 & 0.23 & 0.29 & 0.14 \\
		\multirow{-2}{*}{Carpet} & 0.87 & 0.59 & 0.54 & 0.72 & 0.54 & 0.46 & 0.35 & 0.55 & \textbf{0.88} & 0.57 \\ \hline
		& \textbf{0.88} & 0.83 & 0.04 & 0.02 & - & 0.36 & 0.04 & 0.41 & 0.01 & 0.59 \\
		\multirow{-2}{*}{Grid} & 0.94 & 0.90 & 0.58 & 0.59 & 0.59 & 0.86 & 0.76 & 0.80 & 0.72 & \textbf{0.98} \\ \hline
		& 0.71 & 0.67 & 0.34 & 0.74 & - & 0.09 & 0.12 & 0.31 & \textbf{0.98} & 0.52 \\
		\multirow{-2}{*}{Leather} & 0.78 & 0.75 & 0.64 & 0.87 & 0.73 & 0.65 & 0.64 & 0.77 & \textbf{0.97} & 0.89 \\ \hline
		& 0.04 & 0.23 & 0.08 & 0.14 & - & 0.14 & 0.09 & 0.19 & 0.11 & \textbf{0.23} \\
		\multirow{-2}{*}{Tile} & 0.59 & 0.51 & 0.50 & 0.93 & 0.81 & 0.64 & 0.70 & 0.69 & 0.41 & \textbf{0.97} \\ \hline
		& 0.36 & 0.29 & 0.14 & \textbf{0.47} & - & 0.19 & 0.11 & 0.32 & 0.51 & 0.37 \\
		\multirow{-2}{*}{Wood} & 0.73 & 0.73 & 0.62 & 0.91 & 0.87 & 0.95 & 0.77 & 0.91 & 0.78 & \textbf{0.98} \\ \hline
		& 0.15 & 0.22 & 0.05 & 0.07 & - & 0.09 & 0.11 & 0.13 & - & \textbf{0.43} \\
		\multirow{-2}{*}{Bottle} & 0.93 & 0.86 & 0.86 & 0.78 & 0.86 & 0.76 & 0.73 & 0.82 & - & \textbf{0.99} \\ \hline
		& 0.01 & 0.05 & 0.01 & 0.13 & - & 0.02 & 0.05 & 0.14 & - & \textbf{0.16} \\
		\multirow{-2}{*}{Cable} & 0.82 & \textbf{0.86} & 0.78 & 0.79 & 0.71 & 0.61 & 0.60 & 0.83 & - & 0.70 \\ \hline
		& 0.09 & 0.11 & 0.04 & 0.00 & - & 0.04 & 0.19 & 0.51 & - & \textbf{0.64} \\
		\multirow{-2}{*}{Capsule} & \textbf{0.94} & 0.88 & 0.84 & 0.84 & 0.69 & 0.61 & 0.59 & 0.72 & - & 0.84 \\ \hline
		& 0.00 & 0.41 & 0.02 & 0.00 & - & 0.33 & 0.34 & 0.37 & - & \textbf{0.66} \\
		\multirow{-2}{*}{Hazelnut} & 0.97 & 0.95 & 0.87 & 0.72 & 0.71 & 0.87 & 0.75 & 0.86 & - & \textbf{0.97} \\ \hline
		& 0.01 & \textbf{0.26} & 0.00 & 0.13 & - & 0.04 & 0.01 & 0.18 & - & 0.24 \\
		\multirow{-2}{*}{Metal Nut} & \textbf{0.89} & 0.86 & 0.76 & 0.82 & 0.75 & 0.43 & 0.46 & 0.69 & - & 0.79 \\ \hline
		& 0.07 & 0.25 & 0.17 & 0.00 & - & 0.29 & 0.01 & 0.17 & - & \textbf{0.58} \\
		\multirow{-2}{*}{Pill} & 0.91 & 0.85 & 0.87 & 0.68 & 0.77 & 0.80 & 0.62 & 0.76 & - & \textbf{0.91} \\ \hline
		& 0.03 & \textbf{0.34} & 0.01 & 0.00 & - & 0.17 & 0.02 & 0.24 & - & 0.32 \\
		\multirow{-2}{*}{Screw} & 0.96 & 0.96 & 0.80 & 0.87 & 0.64 & 0.95 & 0.97 & 0.72 & - & \textbf{1.00} \\ \hline
		& 0.08 & 0.51 & 0.07 & 0.00 & - & 0.13 & 0.10 & 0.48 & - & \textbf{0.63} \\
		\multirow{-2}{*}{Toothbrush} & 0.92 & 0.93 & 0.90 & 0.77 & 0.70 & 0.70 & 0.67 & 0.82 & - & \textbf{0.99} \\ \hline
		& 0.01 & 0.22 & 0.08 & 0.03 & - & 0.20 & 0.05 & 0.15 & - & \textbf{0.24} \\
		\multirow{-2}{*}{Transistor} & \textbf{0.90} & 0.86 & 0.80 & 0.66 & 0.65 & 0.72 & 0.78 & 0.79 & - & 0.82 \\ \hline
		& 0.10 & 0.13 & 0.01 & 0.00 & - & 0.05 & 0.04 & 0.21 & - & \textbf{0.34} \\
		\multirow{-2}{*}{Zipper} & 0.88 & 0.77 & 0.78 & 0.76 & 0.74 & 0.63 & 0.60 & 0.84 & - & \textbf{0.90} \\ \hline \hline
		& 0.22 & 0.33 & 0.09 & 0.13 & - & 0.15 & 0.09 & 0.27 & - & \textbf{0.41} \\
		\multirow{-2}{*}{\textbf{Mean}} & 0.87 & 0.82 & 0.74 & 0.78 & 0.72 & 0.71 & 0.66 & 0.77 & - & \textbf{0.89} \\ \hline
	\end{tabular}
	\label{tab:mvtec_results}
\end{table}

\section{Conclusion}


In this work, we propose a novel P-Net for retina image anomaly detection. The motivation of our method is the correlation between structure and texture in healthy retinal images. Our model extracts structure from original images first, and then reconstructs the original images by using both structure information and texture information. At last, we extract the structure from the reconstructed images, and minimizing the difference between the structures extracted from original image and that from reconstructed image. Then we combine the image reconstruction error and structure difference as a measurement for anomaly detection. 
Extensive experiments validate the effectiveness of our approach.

\noindent
\textbf{Acknowledge:} The work was supported by National Key R\&D Program of China (2018AAA0100704), NSFC \#61932020, Guangdong Provincial Key Laboratory (2020B121201001), ShanghaiTech-Megavii Joint Lab, and ShanghaiTech-UnitedImaging Joint Lab.

\clearpage

\bibliographystyle{splncs}
\bibliography{IEEEabrv}

\end{document}